\documentclass[journal,twoside,web]{ieeecolor}
\usepackage{tmi}
\usepackage{cite}
\usepackage{amsmath,amssymb,amsfonts}
\usepackage{algorithmic}
\usepackage{algorithm}
\usepackage{graphicx}
\usepackage{textcomp}
\usepackage{hyperref}

\def\BibTeX{{\rm B\kern-.05em{\sc i\kern-.025em b}\kern-.08em
    T\kern-.1667em\lower.7ex\hbox{E}\kern-.125emX}}
\renewcommand{\vec}[1]{\ensuremath{\boldsymbol{#1}}}

\newcommand{\hvec}[1]{\ensuremath{\Hat{\boldsymbol{#1}}}}

\begin{document}
\title{EMORe: Motion-Robust 5D MRI Reconstruction via Expectation-Maximization-Guided Binning Correction and Outlier Rejection}
\author{Syed M. Arshad, Lee C. Potter, Yingmin Liu, Christopher Crabtree, Matthew S. Tong,  and Rizwan Ahmad
\thanks{This work has been submitted to the IEEE TMI for possible publication. Copyright may be transferred without notice, after which this version may no longer be accessible. This work was supported by the National Heart, Lung, and Blood Institute, Grant/Award Numbers: R01HL135489, R01HL151697. (Corresponding author: Rizwan Ahmad.)}
\thanks{S. M. Arshad and R. Ahmad are with the Department of Biomedical Engineering and the Department of Electrical \& Computer Engineering, The Ohio State University, Columbus, OH 43210 USA (e-mail: arshad.32@osu.edu; ahmad.46@osu.edu). L. C. Potter is with the Department of Electrical \& Computer Engineering, The Ohio State University, Columbus, OH 43210 USA. Y. Liu and C. Crabtree are with the Davis Heart and Lung Research Institute, The Ohio State University Wexner Medical Center, Columbus, OH 43210 USA. M. S. Tong is with the Department of Internal Medicine, The Ohio State University Wexner Medical Center, Columbus, OH 43201 USA.}}

\maketitle

\begin{abstract}
We propose EMORe, an adaptive reconstruction method designed to enhance motion robustness in free-running, free-breathing self-gated 5D cardiac magnetic resonance imaging (MRI). Traditional self-gating-based motion binning for 5D MRI often results in residual motion artifacts due to inaccuracies in cardiac and respiratory signal extraction and sporadic bulk motion, compromising clinical utility. EMORe addresses these issues by integrating adaptive inter-bin correction and explicit outlier rejection within an expectation-maximization (EM) framework, whereby the E-step and M-step are executed alternately until convergence. In the E-step, probabilistic (soft) bin assignments are refined by correcting misassignment of valid data and rejecting motion-corrupted data to a dedicated outlier bin. In the M-step, the image estimate is improved using the refined soft bin assignments. Validation in a simulated 5D MRXCAT phantom demonstrated EMORe's superior performance compared to standard compressed sensing reconstruction, showing significant improvements in peak signal-to-noise ratio, structural similarity index, edge sharpness, and bin assignment accuracy across varying levels of simulated bulk motion. In vivo validation in $13$ volunteers further confirmed EMORe's robustness, significantly enhancing blood-myocardium edge sharpness and reducing motion artifacts compared to compressed sensing, particularly in scenarios with controlled coughing-induced motion. Although EMORe incurs a modest increase in computational complexity, its adaptability and robust handling of bulk motion artifacts significantly enhance the clinical applicability and diagnostic confidence of 5D cardiac MRI.
\end{abstract}

\section{Introduction}
\label{sec:introduction}
\IEEEPARstart{V}{olumetric} cardiovascular magnetic resonance imaging (CMR) allows a comprehensive assessment of the whole heart, capturing the 3D volume in different motion phases. Volumetric imaging circumvents the major limitations of standard 2D CMR, such as breath hold requirement, through-plane motion, misregistration between slices, and the need for precise pre-planning\cite{dyverfeldt20154d}. Most volumetric CMR sequences on commercial scanners acquire the data or several minutes under electrocardiogram (ECG) guidance and prospective respiratory gating using navigator echoes to resolve cardiac and respiratory motion \cite{salerno2017recent}. However, depending on the breathing pattern and the extent of arrhythmia, this approach may lead to unpredictably long acquisition times, sometimes exceeding ten minutes \cite{goo2018nav}. In addition, navigator echoes disrupt the steady-state of magnetization and thus are not compatible with several common CMR pulse sequences \cite{zaitsev2015motion}. As an alternative to navigator echoes, respiratory bellows \cite{yuan2000bellow} can be used for prospective or retrospective gating, but their use has been limited due to inconsistent performance \cite{santelli2011nobellow}. 
More recently, there has been increased interest in free-running self-gated (SG) volumetric imaging (FRV-CMR) \cite{feng2016sg,di2019sg,ma2020sg,kustner2021sg,4dflow_pruitt2021}, in which data are collected continuously for several minutes without the guidance of navigator echoes, respiratory bellows, or even ECG. Furthermore, FRV-CMR offers flexibility in retrospectively controlling temporal resolution by selecting the number of cardiac and respiratory bins after the scan \cite{holtackers2024flex}.

In SG acquisition \cite{larson2004sg_orig}, a readout passing through the center of the k-space is periodically sampled. The temporal changes in this readout are attributed to physiological motion. A common approach for extracting respiratory and cardiac signals from these repeatedly sampled readouts relies on blind source separation. These surrogate motion signals are then employed to assign the rest of the k-space data into motion bins, leading to motion-resolved volumetric CMR \cite{feng20185d}. There are various approaches for retrospectively binning the cardiorespiratory data, according to clinical application and computational constraints. The first approach is respiratory-motion-compensated and cardiac-motion-resolved imaging \cite{liu20103dcine,usman20173dcine,moghari20183dcine}; this is achieved by keeping and processing the k-space data from only one respiratory phase. The limitation of this approach is that data efficiency is $50\%$ or lower because the data from only one respiratory phase are utilized. The second approach is respiratory-motion-corrected and cardiac-motion-resolved imaging \cite{stehning2005cor3dcine,bhat2011cor3dcine,ingle2014cor3dcine}; this is achieved by registering or deforming all other respiratory phases to one reference phase during the reconstruction. Although this approach provides $100\%$ data efficiency, it requires estimation of high-quality deformation fields, which is a difficult task. Moreover, this approach averages the effect of respiratory phases on cardiac function. The third and more recent approach is cardiorespiratory-motion-resolved imaging, commonly known as 5D MRI \cite{wu20125d,feng20185d,sieber20255d}. There has been increased interest in 5D MRI among researchers and clinicians due to its $100\%$ imaging efficiency, ability to resolve respiratory effects on cardiac function \cite{weiss2024resp}, and introduction of a respiratory dimension for enhanced temporal regularization \cite{feng2016xd,ma2020sg}. Nevertheless, the use of SG in all three approaches assumes the availability of reliable cardiac and respiratory motion signals, which are often extracted using blind source separation techniques. While these methods do not require perfectly periodic motion, irregularities such as patient movement, beat-to-beat variability, or inconsistent breathing can affect the accuracy of signal extraction and bin assignment \cite{arshad2024CORe}. Consequently, in all three binning approaches, motion suppression may be degraded by inaccuracies in the extracted signals and by motion outliers from involuntary bulk motion such as deep breaths, coughing, sneezing, or twitches. These inaccuracies can lead to incorrect assignment of k-space data to motion bins, resulting in residual motion artifacts and image blurring. Although Pilot Tone (PT) \cite{ludwig2021pt} provides an alternative to SG, the extraction of physiological motion from PT faces similar challenges.

Recent works on FRV-CMR reconstruction based on compressive recovery\cite{feng2016xd,4dflow_pruitt2021,roy20245d} or deep learning models\cite{kustner2020cinenet} do not account for the discussed imperfections in motion estimation. We posit that residual motion from imperfect retrospective binning can arise from two categories: (i) valid data assigned to a wrong motion bin due to inaccurate self-gating signal and (ii) inclusion of outliers from, for example, bulk motion. For outlier rejection, static imaging protocols often utilize quantitative motion assessment techniques, such as correlation with reference images or temporal consistency, to identify and exclude motion-corrupted data using a fixed threshold\cite{niethammer2007staticoutlier1,dolui2017staticoutlier2,patel2014staticoutlier3}. However, these approaches lack adaptability to the dynamic spatiotemporal correlations inherent in CMR. Recent advances in outlier rejection in dynamic imaging leverage physics-guided group sparsity \cite{arshad2024CORe} or use of expectation-maximization (EM) \cite{em_mclachlan2008} to determine inliers and outliers using a mixture model of the two classes \cite{van2018emoutlier}. However, these outlier rejection techniques suffer from the major limitation of discarding incorrectly binned, yet valid, data as outliers and effectively increasing the acceleration rate, as they lack the flexibility to correct inaccurate bin assignments. There is recent work on intra-bin motion correction \cite{roy20245d} proposing rigid motion correction within a respiratory bin to reduce respiratory blurring in 5D MRI. While their method improves sharpness within respiratory bins, it also assumes perfect signal extraction from self-gating and accurate initial bin assignments.

Our proposed reconstruction framework, expectation-maximization (EM)-guided binning correction and outlier rejection (EMORe), integrates adaptive outlier rejection and binning refinement within a single iterative reconstruction framework using EM. Unlike previous works, EMORe does not assume self-gating-based binning as final, but rather leverages intra-bin inconsistencies to iteratively reassign data readouts to their correct motion bins or to reject a readout to an outlier bin, with corresponding improvement in image estimates. 
This study extends the preliminary results on a synthetic 2D cine phantom presented at the SCMR 2025 Annual Scientific Sessions~\cite{arshad2025emore}. In this work, we present the detailed implementation of the EMORe algorithm for free-breathing volumetric CMR. We validate its performance against standard compressed sensing using a 5D MRXCAT phantom. We also assess its effectiveness in 5D MRI reconstruction using data collected from human subjects.

\section{Methods}
\subsection{EM–Guided Binning Correction and Outlier Rejection (EMORe)}\label{sec:emore}
In FRV-CMR, the acquired $N$ data readouts are retrospectively assigned to $K$ target cardiorespiratory motion bins using SG. However, as discussed in Section~\ref{sec:introduction}, the initial SG-based bin assignments of $N$
acquired readouts to $K$ motion states, or bins, is inherently imperfect.
Let $g_n \in \{1, \dots, K\}$ denote the SG-based bin assignment for the $n^{\text{th}}$ readout.
To mitigate inaccurate data binning, the proposed EMORe framework leverages the EM algorithm by treating the true motion bin assignments as latent random variables, represented by $r_n\in \{1, \dots, K, K+1\}$, where the additional $(K+1)^{\text{th}}$ bin corresponds to an outlier bin for discarding readouts corrupted by bulk motion that do not belong to any valid motion bin $\{1, \dots, K\}$. The algorithm learns the probabilities that a readout belongs to any bin, given the k-space data and the current image estimates; a readout then participates in image formation for any motion state bin weighted by these probabilities.

\begin{figure*}[!t]
\centerline{\includegraphics[width=6in]{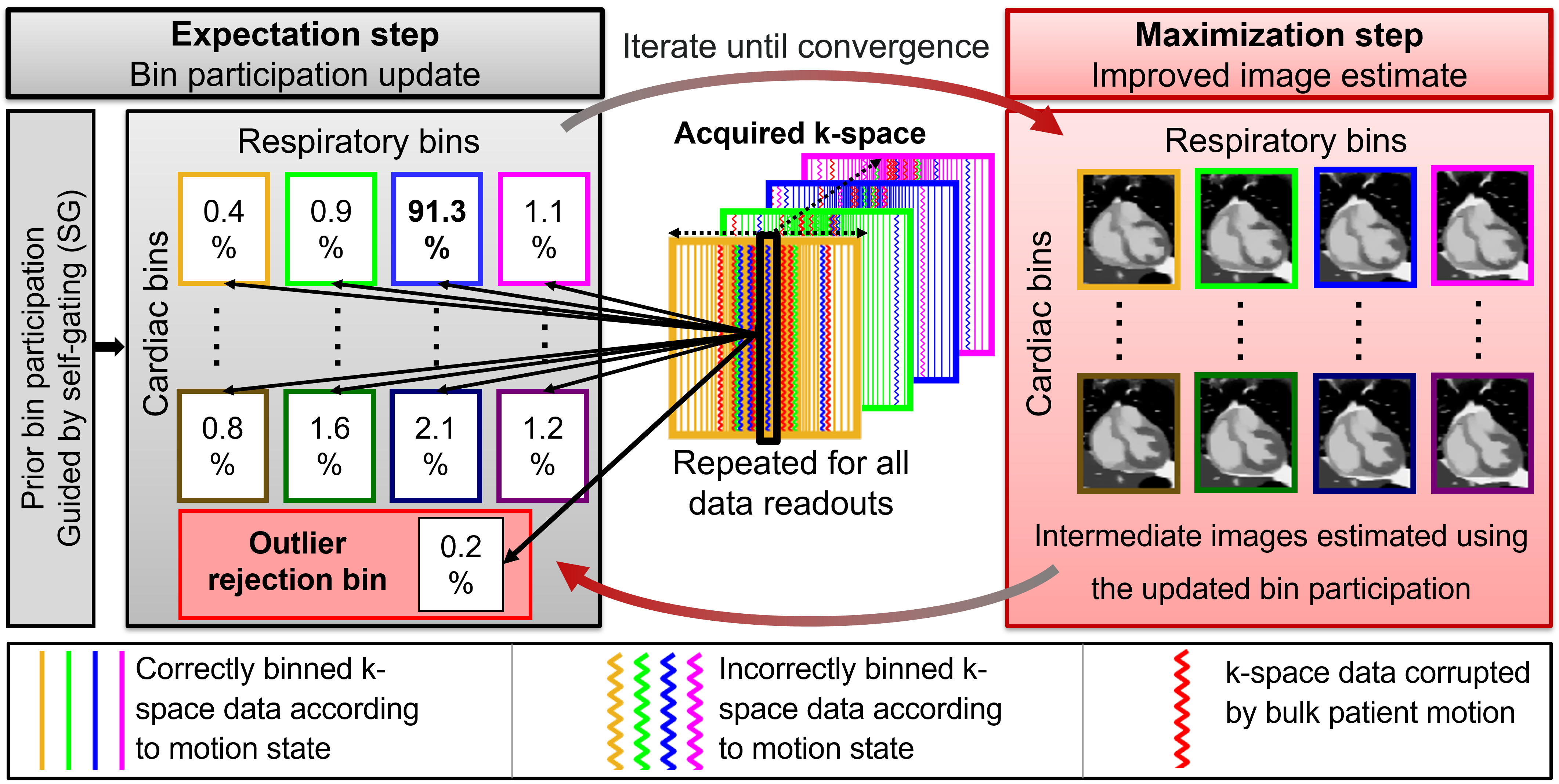}}
\caption{Schematic overview of the proposed EMORe framework. (Left) In the E-step \eqref{eq:estep}, we refine bin participation of readouts to valid motion bins and an outlier bin, given the prior bin participation and current image estimate. (Right) In the M-step \eqref{eq:mstep}, we improve the image estimate using the refined bin participation. Both steps are repeated until convergence, resulting in motion-compensated images.}
\label{fig:em}
\end{figure*}

The EMORe algorithm iteratively performs two steps until convergence: the E-step and the M-step as shown in Fig.~\ref{fig:em}. In the E-step, EMORe refines the probabilistic (soft) participation of acquired readouts across the valid motion bins and outlier bin based on the current estimates of the images for all motion states. Subsequently, the M-step updates the image estimates utilizing the refined bin participation obtained from the E-step. By integrating explicit inter-bin correction and outlier rejection within this EM framework, EMORe enhances robustness against incorrect binning, bulk motion and other types of data corruption, thereby significantly improving the quality of reconstructed images. Detailed implementations of the E-step and M-step in the EMORe framework are described below.

\subsubsection{E-step}\label{sec:estep}
In the E-step of the EMORe framework, we use the current estimate of the images and the acquired data to update the probabilistic (soft) participation weights for each readout across all $K+1$ motion bins. The participation weight of the $n^{th}$ readout to $k^{th}$ bin at iteration $t$, denoted $w_{(n,k)}^{(t)}$, represents the posterior probability of the readout belonging to the bin. Via Bayes' theorem these weights are given by:
\begin{subequations}\label{eq:estep}
\begin{align*}
w_{(n,k)}^{(t)} &= p(r_n=k|\vec{y}_n,\hvec{x}^{(t-1)}), \notag\\
                &= \frac{p(\vec{y}_n|\hvec{x}^{(t-1)},r_n=k)\cdot p(r_n=k|\hvec{x}^{(t-1)})}
{\sum_{j=1}^{K+1} p(\vec{y}_n|\hvec{x}^{(t-1)},r_n=j)\cdot p(r_n=j|\hvec{x}^{(t-1)})}.
\end{align*}
Assuming that the prior probability of the $n^{th}$ readout belonging to $k^{th}$ bin, $p(r_n=k)=\theta_{(n,k)}$, is independent of the image estimate,
\begin{align}
w_{(n,k)}^{(t)} &= \frac{p(\vec{y}_n|\hvec{x}^{(t-1)},r_n=k)\cdot\theta_{(n,k)}}{\sum_{j=1}^{K+1} p(\vec{y}_n|\hvec{x}^{(t-1)},r_n=j)\cdot\theta_{(n,j)}}.\label{eq:w_gen}
\end{align}
\text{For valid motion bins $(k = 1, \dots, K)$,}
\begin{equation}
p(\vec{y}_n|\hvec{x}^{(t-1)},r_n=k) = 
\frac{\exp\left(-\frac{1}{L\sigma^2}\|\vec{A}_{(n,k)}\hvec{x}^{(t-1)}-\vec{y}_n\|_2^2\right)}{\pi\sigma^2};\label{eq:w_k}
\end{equation}
\text{and, for the outlier motion bin $(k = K+1)$,}
\begin{equation}
p(\vec{y}_n|\hvec{x}^{(t-1)},r_n=K+1) =
\frac{\exp\left(-\frac{\tau^2}{\sigma^2}\right)}{\pi\sigma^2}.\label{eq:w_out}
\end{equation}
\end{subequations}
Here, $\vec{A}_{(n,k)}\in \mathbb{C}^{L\times M}$ is the multi-coil forward operator that incorporates pixel-wise multiplication with sensitivity maps, Fourier transform, and selection of the $n^{th}$ readout assigned to the $k^{th}$ motion state bin. Additionally, $\sigma$ is the standard deviation of the circularly symmetric Gaussian noise in the measured data, $L$ is the collective length of a readout from all coils, and $M$ is the number of voxels in the volumetric image for each motion state. Given that the duration of a single readout is negligible relative to physiological motion, we assume motion affects the entire readout vector $\vec{y}_n$ and normalize the squared residual norm by $L$ in \eqref{eq:w_k} to estimate probabilities per readout. The parameter $\tau$ determines the degree of outlier data rejection: a smaller value of $\tau$ corresponds to more aggressive outlier rejection at the potential cost of discarding valid data. The choice of bin prior probabilities, $\theta_{(n,k)}$, is detailed in Section~\ref{sec:algo}.
\subsubsection{M-step}\label{sec:mstep}
In the M-step of the EMORe framework, we use the updated readout participation weights $w_{(n,k)}^{(t)}$ to re-estimate the images $\hvec{x}^{(t)}$. For additive white Gaussian measurement noise, this step minimizes the negative log-posterior probability of the images $\vec{x}$ given the readout measurements $\vec{y}_n$, updated weights $w_{(n,k)}^{(t)}$, and a regularizing prior proportional to $e^{-\mathcal{R}(\vec{x})}$. The corresponding optimization problem is formulated as:
\begin{align}
\hvec{x}^{(t)} &=\arg\min_{\vec{x}} \sum_{n=1}^{N} \sum_{k=1}^K\left[-w_{(n,k)}^{(t)}\log p(\vec{y}_n|r_n=k,\vec{x})\right]+\mathcal{R}(\vec{x})\notag\\
&= \arg\min_{\vec{x}} \sum_{n=1}^{N} \sum_{k=1}^{K} 
\left[\frac{w_{(n,k)}^{(t)}}{L\sigma^2}\|\vec{A}_{(n,k)}\vec{x}-\vec{y}_n\|_2^2\right]+ \notag \\
& \hspace*{18mm} \lambda_s \|\nabla_{s} \vec{x}\|_1 + \lambda_c \|\nabla_{c} \vec{x}\|_1 + \lambda_r \|\nabla_{r} \vec{x}\|_1.
\label{eq:mstep}
\end{align}
where $\|\nabla_{s}\vec{x}\|_1$, $\|\nabla_{c}\vec{x}\|_1$, and $\|\nabla_{r}\vec{x}\|_1$ denote the anisotropic total variation (TV) operator enforcing smoothness along three spatial dimensions, the cardiac temporal dimension, and the respiratory temporal dimension, respectively. The hyperparameters $\lambda_s$, $\lambda_c$, and $\lambda_r$ control the strengths of regularization in their respective dimensions. The optimization problem in \eqref{eq:mstep} is solved using the alternating direction method of multipliers (ADMM)\cite{boyd2011admm}. In \eqref{eq:mstep}, the data fidelity term is normalized by $L$ to maintain consistency with \eqref{eq:w_k}.

\subsubsection{Algorithm parameters}\label{sec:algo}
The complete EMORe algorithm for FRV-CMR reconstruction is presented in Algorithm~\ref{alg:EM_FRV}. The EM algorithm is known to be sensitive to initialization\cite{em_mclachlan2008}, so we initialize $\hvec{x}^{(0)}$ by partially solving \eqref{eq:mstep} for $I_1$ ADMM iterations, using initial readout participation weights derived from SG-based bin assignment as follows,
\begin{equation}
    w_{(n,k)}^{(0)} =
    \begin{cases}
        1, & \text{if } k = g_n, \\[3pt] 
        0, & \text{otherwise}.
    \end{cases}
    \quad \forall\, n,k.
    \label{eq:w_sg}
\end{equation}
While this SG-based assignment is imperfect, it provides adequate initialization for convergence of the algorithm to a reasonable local minimum.

In the high-dimensional inverse problem of FRV-CMR, the EM algorithm can become unstable\cite{kadir2014high}, especially when the total number data bins $K$ is large (e.g., $K=80$, for $20$ cardiac $\times$ $4$ respiratory bins). To mitigate this instability, we leverage a bin-assignment prior $\theta_{(n,k)}$, where $\sum_{k=1}^{K+1} \theta_{(n,k)}=1$ for every $n$. We define an informative prior guided by SG-based initial assignments as follows:
\begin{equation}
\theta_{(n,k)} =
\begin{cases}
    \alpha_g, & \text{if } k = g_n, \\ 
    \alpha_o, & \text{if } k = K+1, \\ 
     (1-\alpha_g-\alpha_o)/(K-1), & \text{otherwise}.
\end{cases}
\label{eq:theta}
\end{equation}
The hyperparameters $\alpha_g$ and $\alpha_o$ are probabilities, chosen to balance the effectiveness of residual motion compensation against algorithm convergence stability. Specifically, a higher value of $\alpha_g$ increases reliance on the initial SG-based binning, enhancing stability but limiting motion correction. Whereas, a higher value of $\alpha_o$ promotes aggressive rejection of outliers, at the potential cost of discarding valid data.

The M-step \eqref{eq:mstep} presents
a large-scale optimization problem due to the large size of each volumetric image.
Therefore, obtaining a closed-form solution or fully solving the M-step problem through iterative procedures is computationally expensive and impractical. To address this, we employ a generalized EM \cite{em_mclachlan2008} approach, where the M-step is approximately solved via $I_2$ ADMM iterations.
The stopping criterion of the EMORe algorithm is either the maximum allowed number of EM iterations, denoted by $J$, or a threshold $\eta$
on the normalized squared image difference between consecutive EM iterations -- whichever is achieved first.
Algorithm hyperparameters are given in Table~\ref{tab:params}. These values were optimized using one digital dataset from the phantom study. 

\begin{algorithm}
\caption{EMORe reconstruction}
\label{alg:EM_FRV} 
\begin{algorithmic}[1]
\STATE \textbf{Initialize} $\hvec{x}^{(0)}$ using $w_{(n,k)}^{(0)}$ in \eqref{eq:mstep} and performing $I_1$ ADMM iterations. \label{step:init_x}
\STATE \textbf{Set} $t \gets 1$ \quad \textit{(Initialize iteration counter)} \label{step:init_t}
\REPEAT
    \STATE \textbf{E-step}: Update $w_{(n,k)}^{(t)}$ using $\hvec{x}^{(t-1)}$ in \eqref{eq:estep}. \label{step:update_w}

    \STATE \textbf{M-step}: Estimate $\hvec{x}^{(t)}$ by using $w_{(n,k)}^{(t)}$ in \eqref{eq:mstep} and performing $I_2$ ADMM iterations. \label{step:update_x}
    \STATE $t \gets t + 1$ \quad \textit{(Increment iteration counter)} \label{step:increment_t}
\UNTIL{$\|\hvec{x}^{(t)}-\hvec{x}^{(t-1)}\|_2^2/\|\hvec{x}^{(t-1)}\|_2^2 < \eta \land t \geq J$} \label{step:converg}
\RETURN $\hvec{x}^{(t)}$ 
\end{algorithmic}
\end{algorithm}
\subsection{Image comparison \& evaluation}
Compressed sensing (CS)\cite{cs_lustig2008}, the predominant FRV-CMR reconstruction method and widely used in both clinical and research settings, was employed as the standard for comparison with EMORe. The CS image estimate, $\hat{\vec{x}}_{CS}$, was obtained by solving a regularized weighted least-squares problem in \eqref{eq:mstep} using binary participation weights derived from the initial SG-based bin assignments $w_{(n,k)}^{(0)}$ \eqref{eq:w_sg} for all iterations, $t$.

Similar to the EMORe, the ADMM algorithm was employed to solve the optimization problem in the CS reconstruction, with iterations terminating upon reaching a maximum count $J$ or when the normalized squared difference between consecutive image iterates fell below a threshold $\eta$.

In phantom studies, where a reference image was available, we compared the quality of EMORe and CS images using the structural similarity index (SSIM$\uparrow$)\cite{wang2004ssim} and peak signal-to-noise ratio (PSNR$\uparrow$), defined as $20\log_{10} \left( \frac{\max(\vec{x})}{\|\hat{\vec{x}} - \vec{x}\|_2 / \sqrt{N}} \right)$ (dB).  Additionally, we measured blood-myocardium edge sharpness of all reconstructed images using edge sharpness assessment\cite{edge_ahmad2015} presented for MRI. The edge sharpness ($\uparrow$) was quantified as the mean slope of sigmoid functions fitted to pixel intensity profiles crossing the blood-myocardium boundaries, where residual motion artifacts typically introduce blurring. Higher slope values indicate improved edge sharpness and clearer delineation of anatomical structures. In our 5D MRI data, a single blood–myocardium edge sharpness value was obtained by averaging estimates across all four respiratory states, measured on a fixed sagittal cardiac slice where the blood pool was fully enclosed by the myocardium. Furthermore, in studies where the true bin assignments were available, we assessed the improvement in bin participation of data using EMORe by comparing initial and final Brier score\cite{brier1950brier}, which evaluates the mean squared error of probabilistic bin assignments. The Brier score ($\downarrow$) at EMORe iteration $t$ is defined as $\frac{1}{N} \sum_{n=1}^{N} \sum_{k=1}^{K} (\Tilde{w}_{(n,k)} - w_{(n,k)}^{(t)})^2$, where $\Tilde{w}_{(n,k)}$ represents the true bin participation of readout $n$ for bin $k$. In the absence of reference images for the in vivo study, two expert readers independently and blindly evaluated EMORe–CS cine slice pairs for perceived noise and artifacts quality using a 5-point Likert scale (5: excellent, 4: good, 3: fair, 2: poor, 1: non-diagnostic).
\subsection{Experimental studies}
To evaluate the robustness of the proposed EMORe reconstruction framework against uncompensated motion in 5D MRI, we performed phantom and in vivo studies comparing EMORe to conventional CS reconstruction.
\begin{table}[!t]
\caption{Imaging and Reconstruction Parameters for Phantom and In Vivo 5D MRI Studies}
\label{tab:params}
\centering
\setlength{\tabcolsep}{4pt}
\renewcommand{\arraystretch}{1.2}
\begin{tabular}{|p{0.41\columnwidth}|p{0.25\columnwidth}|p{0.25\columnwidth}|}
\hline
\textbf{Parameter} & \textbf{Phantom Study} & \textbf{In Vivo Study} \\
\hline
\hline
\multicolumn{3}{|c|}{\textbf{Imaging parameters}} \\
\hline
\hline
Number of Datasets & $50$ & $13$ \\
\hline
Spatial Resolution (mm) & $2 \times 2 \times 2$ & $1.5$--$2.1 \times 1.5$--$2.1 \times 2.0$--$2.8$ \\
\hline
Matrix Size & $90 \times 82 \times 76$ & $96 \times 144 \times 80$ \\
\hline
Acceleration Rate $R$ & $7.3$ & $9.3$--$9.8$ \\
\hline
Sex Distribution, Male / Female & $20\,/\,30$ & $7\,/\,6$ \\
\hline
Age Range (Mean), years & -- & $20$--$68$ ($36$) \\
\hline
BMI Range (Mean), kg/m\textsuperscript{2} & -- & $22$--$37$ ($28$) \\
\hline
TE / TR (ms) & $1.2 \,/\, 4.0$ & $1.2 \,/\, 3.0$--$3.2$ \\
\hline
Sampling Pattern & 3D Cartesian & 3D Cartesian \\
\hline
Scan Time (min) & $5$ & $\sim5$ \\
\hline
\hline
\multicolumn{3}{|c|}{\textbf{Reconstruction parameters}} \\
\hline\hline
Regularization $\lambda_s$, $\lambda_c$, $\lambda_r$
  & \multicolumn{2}{p{0.5\columnwidth}|}{$2\times10^{-2},\ 10\times10^{-2},\ 6\times10^{-2}$} \\
\hline
Convergence threshold $\eta$
  & \multicolumn{2}{p{0.5\columnwidth}|}{$10^{-4}$} \\
\hline
Outlier threshold $\tau$
  & \multicolumn{2}{p{0.5\columnwidth}|}{$3\sigma$} \\
\hline
Prior $\alpha_{g}$, $\alpha_{o}$
  & \multicolumn{2}{p{0.5\columnwidth}|}{$0.85,\ 0.05$} \\
\hline
Inner iterations $I_1$, $I_2$
  & \multicolumn{2}{p{0.5\columnwidth}|}{$10,\ 4$} \\
\hline
Max outer iterations $J$
  & \multicolumn{2}{p{0.5\columnwidth}|}{$60$} \\
\hline
\end{tabular}
\end{table}

\subsubsection{5D MRXCAT phantom study}\label{sec:5DMRX}
We simulated free-breathing self-gated 5D MRI scans using the MRXCAT phantom~\cite{mrxcat_wissmann2014} from five digital subjects. Each subject comprised four respiratory phases (end-expiratory to end-inspiratory) and 20 cardiac phases spanning the full cardiac cycle, resulting in $4 \times 20 = 80$ cardiorespiratory phases. Respiratory periods ranged from $3.25$ to $4.75$ s across the subjects, with a random variation of $0$ to $1$ s introduced between consecutive cycles to mimic irregular breathing. Cardiac activity of the digital subjects was simulated with heart rates ranging between $62$ and $83$ beats per minute, with additional beat-to-beat variations of $0$ to $160$ ms in consecutive R–R intervals.

The k-space data were sampled using a pseudo-random Cartesian trajectory with self-gating~\cite{joshi2022sampling}, acquiring $75,000$ readouts over a 5-minute scan with a repetition time (TR) of $4$ ms. Coil sensitivity maps for an 8-channel array were generated using the Biot–Savart law, with coils positioned evenly in anterior and posterior planes. Complex circularly symmetric Gaussian noise was added to achieve a SNR of $30$ dB. Additional imaging parameters are detailed in Table~\ref{tab:params}.

To simulate sporadic bulk-motion, seven distinct outlier motion states were generated per subject in addition to the reference (resting) state. These included rigid-body translations of $\pm20$~mm along the superior–inferior axis and rotations of $\pm10^\circ$ about the anterior–posterior and superior-inferior axes, as well as a rotation of $-10^\circ$ about the left–right axis. A positive rotation about the left–right axis was omitted, reflecting the low likelihood of backward head-tilting motion in typical patient scanning. 
For each of five digital subjects, ten datasets were generated by introducing varying levels of bulk-motion corruption: $0\%$, $5\%$, $10\%$, $15\%$, $20\%$, $30\%$, $40\%$, $50\%$, $60\%$, and $70\%$. Within each of these $50$ simulated scans, the bulk-motion corruption was distributed into ten different episodes. 
These bulk-motion episodes ranged in duration from approximately $1.5$ s (in the $5\%$ outlier scenario) to $21$ s (in the extreme $70\%$ outlier scenario). In each bulk-motion episode, one of the seven outlier motion states was randomly selected to replace the corresponding reference k-space data.

During data acquisition simulation, the SG signal was sampled every ten readouts in the superior-inferior direction. In the pre-preprocessing of the scanned data, the SG signal was reorganized into a Casorati matrix, and two band-pass filters with passbands of $0.1$–$0.5$ Hz (respiratory) and $0.5$–$3$ Hz (cardiac) were applied along the temporal dimension. Subsequently, principal component analysis (PCA) and independent component analysis (ICA) were performed on the filtered Casorati matrix to estimate respiratory and cardiac surrogate signals \cite{4dflow_pruitt2021,chen2024cardiac}. Based on these signals, k-space data were binned into four respiratory bins of equal data efficiency, each further divided into 20 cardiac bins by partitioning the R-R interval into equal-duration phases, resulting in $K=80$ cardiorespiratory bins.

The SG-based binary bin assignments were used to initialize the participation weights $w_{(n,k)}^{(0)}$ and prior $\theta_{(n,k)}$ using \eqref{eq:w_sg} and \eqref{eq:theta}, respectively. Due to cardiac and respiratory variability, the binning is imperfect, even in the case of $0\%$ outlier corruption fraction.
An initial image estimate $\hvec{x}^{(0)}$ was obtained by partially solving \eqref{eq:mstep} for $I_1$ ADMM iterations. Subsequently, the EMORe reconstruction iteratively performed the E-step and M-step to refine bin assignments, reject outlier data, and enhance image quality until convergence criteria were satisfied. For comparison, CS reconstructions were performed for all simulated datasets.

Image quality of EMORe and CS reconstructions was quantitatively assessed across all phantom experiments using PSNR, SSIM, and edge sharpness, using the true phantom images as reference. Additionally, improvement in bin assignments achieved by EMORe was evaluated against the initial SG-based binning employed in CS using the Brier score.   

\subsubsection{In vivo 5D MRI study}\label{sec:5DMRI}
We compared EMORe and CS reconstructions using in vivo 5D MRI data acquired from $13$ healthy volunteers in a study approved by the institutional review board. Written informed consent was obtained from all participants prior to imaging. Imaging was performed on a 3T clinical scanner (MAGNETOM Vida, Siemens Healthcare, Erlangen, Germany) equipped with a 48-channel receiver coil. Ferumoxytol-enhanced scans were obtained using a free-running, free-breathing acquisition with a fixed duration of approximately 5 minutes. Data were sampled using a pseudo-random Cartesian trajectory with SG as in Section~\ref{sec:5DMRX}. Detailed imaging parameters are summarized in Table~\ref{tab:params}. To evaluate the robustness of EMORe in handling controlled bulk motion, three volunteers were instructed to simulate coughing during the final $30$ seconds of their respective scans.

Acquired k-space data were retrospectively sorted into 80 cardiorespiratory bins using blind-source-separation, following methods outlined in Section~\ref{sec:5DMRX}. For computational efficiency, coil compression via singular value decomposition was applied, reducing the original 30 acquisition coils to 8 virtual coils~\cite{coilbuehrer2007}. Coil sensitivity maps were estimated using the approach by Walsh et al.\cite{mapwalsh2000}. Subsequently, both EMORe and CS methods were applied to reconstruct 5D MRI images, using frameworks identical to those described for the phantom study (Section~\ref{sec:5DMRX}).

Due to the lack of a ground truth reference for in vivo data, quantitative image assessment was limited to blood-myocardium edge sharpness. For qualitative evaluation, blind scoring of artifacts was performed on EMORe and CS sagittal cine slice pairs from end-expiratory and end-inspiratory phases of each volunteer's reconstruction, yielding a total of 26 scored cine image pairs.

\subsection{Implementation details}
EMORe and CS reconstructions were implemented in MATLAB (Mathworks, Natick, MA, USA) and executed on an NVIDIA H100 GPU at the Ohio Supercomputer Center. The average reconstruction times were approximately $23.7$ minutes for EMORe and $19.1$ minutes for CS in the phantom study, and $58.0$ minutes for EMORe and $49.8$ minutes for CS in the in vivo study. The source code and one sample dataset are publicly available at github.com/OSU-MR/Motion-Robust-5D-MRI-EMORe.

\section{Results}
\subsection{5D MRI phantom Study}
\begin{figure}[!t]
\centerline{\includegraphics[width=\columnwidth]{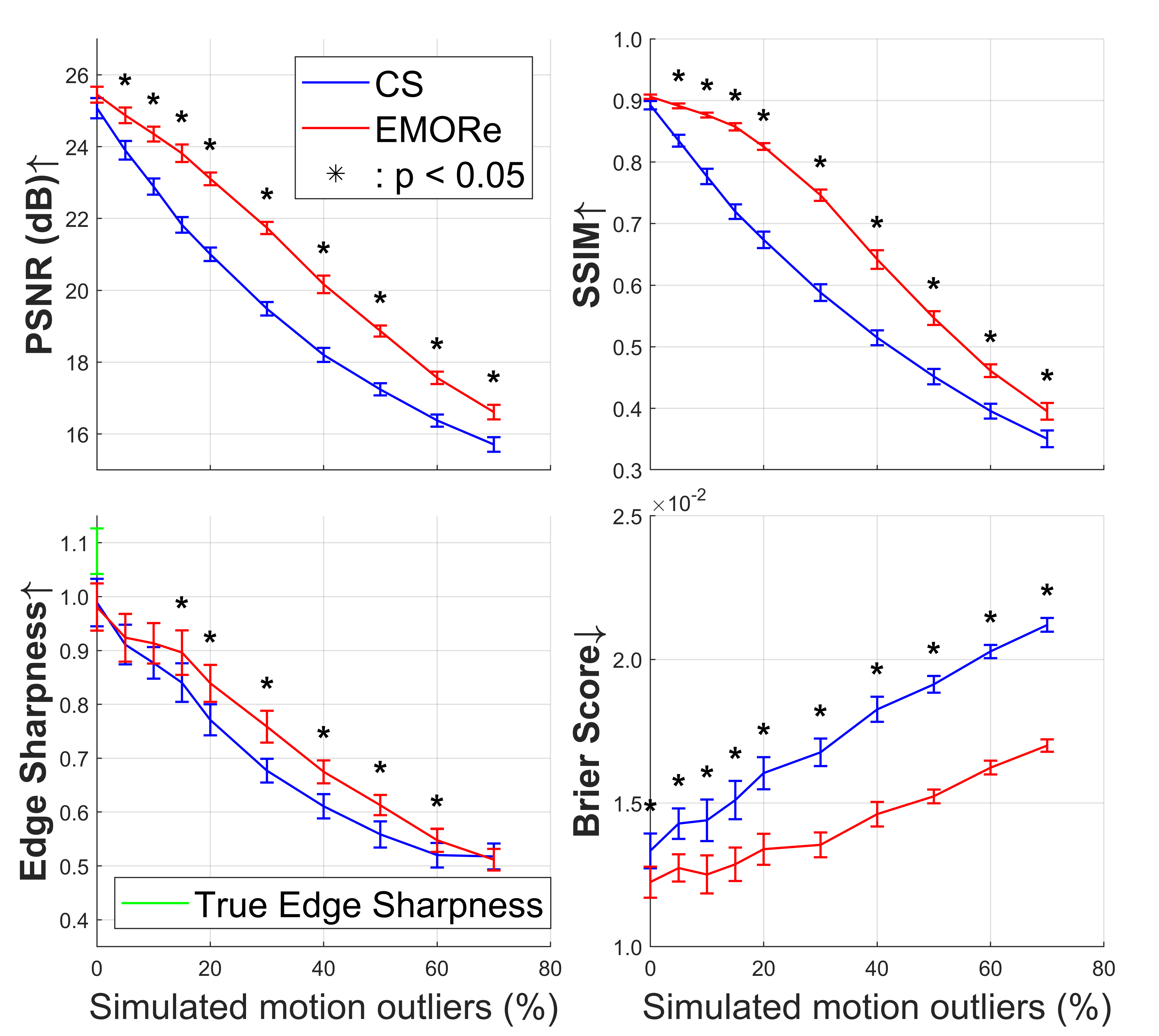}}
\caption{Quantitative comparison between EMORe (red) and CS (blue) reconstructions for 5D MRXCAT phantom study across varying levels of simulated motion outliers ($0$--$70\%$). Metrics shown include PSNR (dB), SSIM, edge sharpness, and Brier score, averaged across five digital subjects. Error bars represent standard error of the mean. Asterisks indicate statistical significance ($p < 0.05$) using a paired $t$-test across five subjects at each outlier level. The edge sharpness (green) for the true phantom with no outliers is also shown for comparison.}
\label{fig:phantom_study}
\end{figure}

The quantitative comparisons between EMORe and CS reconstructions across five MRXCAT digital subjects at ten simulated outlier corruption levels are presented in Fig.~\ref{fig:phantom_study}. Metrics reported include PSNR, SSIM, edge sharpness, and Brier score, averaged across subjects. Statistical significance was assessed using a paired $t$-test across subjects at each outlier level, with significance indicated for $p < 0.05$.

\begin{figure*}[!t]
\centerline{\includegraphics[width=5in]{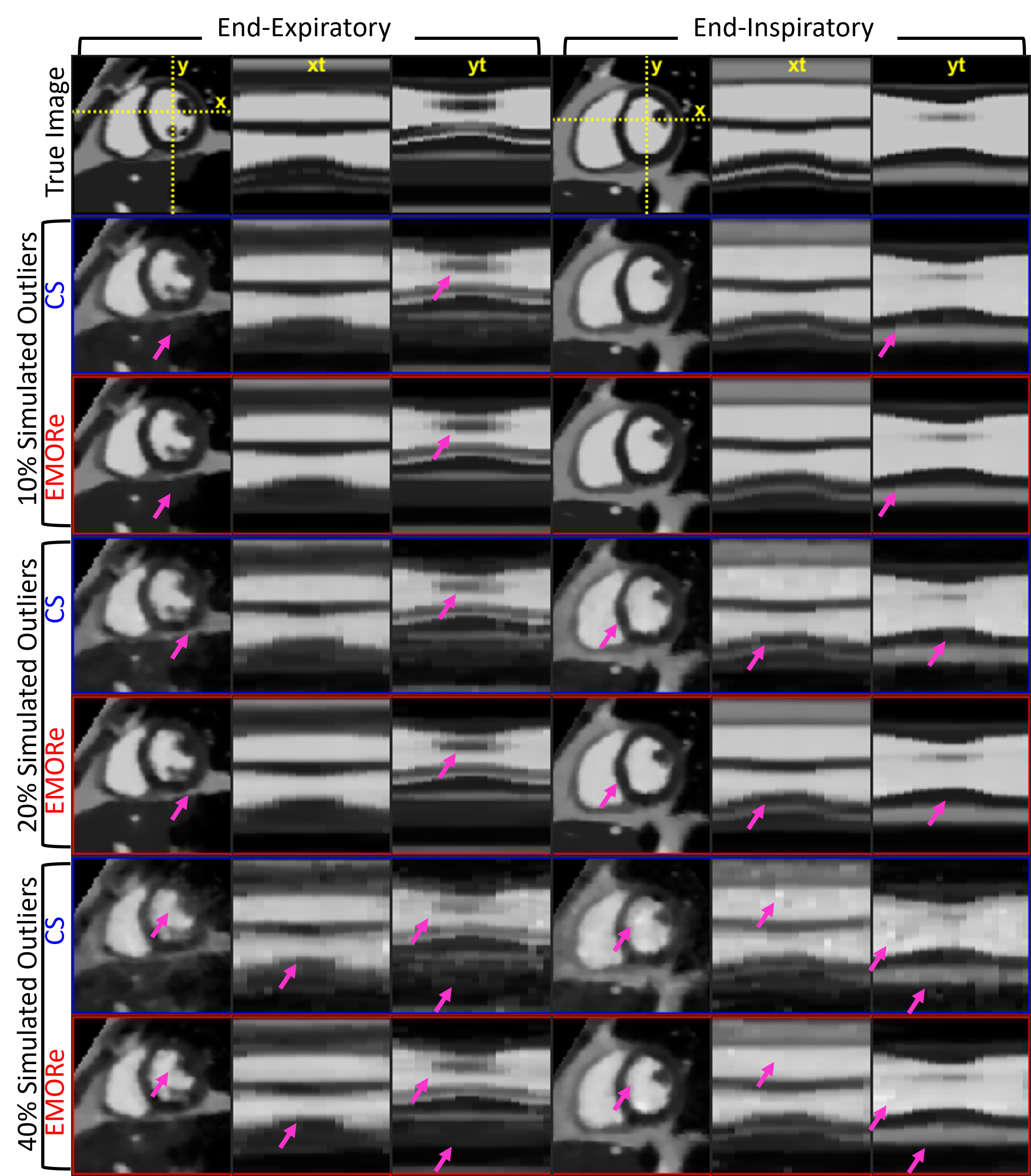}}
\caption{Representative short-axis slices at end-expiratory and end-inspiratory states extracted from 5D MRXCAT reconstructions using CS and EMORe under $10\%$, $20\%$, and $40\%$ simulated motion outlier levels. Each group shows the static frame, and the corresponding temporal profiles along the $x$–$t$ and $y$–$t$ dimensions, extracted at the indicated spatial positions (yellow crosshairs). Arrows highlight motion artifacts and blurring in CS images that are visibly reduced in EMORe reconstructions. }
\label{fig:phantom_fig}
\end{figure*}

Representative end-expiratory and end-inspiratory short-axis cine slices from the ground truth, CS, and EMORe reconstructions are shown in Fig.~\ref{fig:phantom_fig} for three scenarios: $10\%$, $20\%$, and $40\%$ simulated outliers. Corresponding temporal profiles along the $x$–$t$ and $y$–$t$ dimensions are also visualized to facilitate qualitative comparison of temporal consistency and boundary definition. A video\footnote{The supplementary video contains 20 cardiac frames from both end-expiratory and end-inspiratory slices shown in Fig.~\ref{fig:phantom_fig} and Fig.~\ref{fig:real_fig}, played 10 times at 10 frames per second. The file is provided in MP4 format.\label{fn:video}} containing all the cine slices presented in Fig.~\ref{fig:phantom_fig} is available under the “Ancillary files”.

\begin{figure*}[!t]
\centerline{\includegraphics[width=\textwidth]{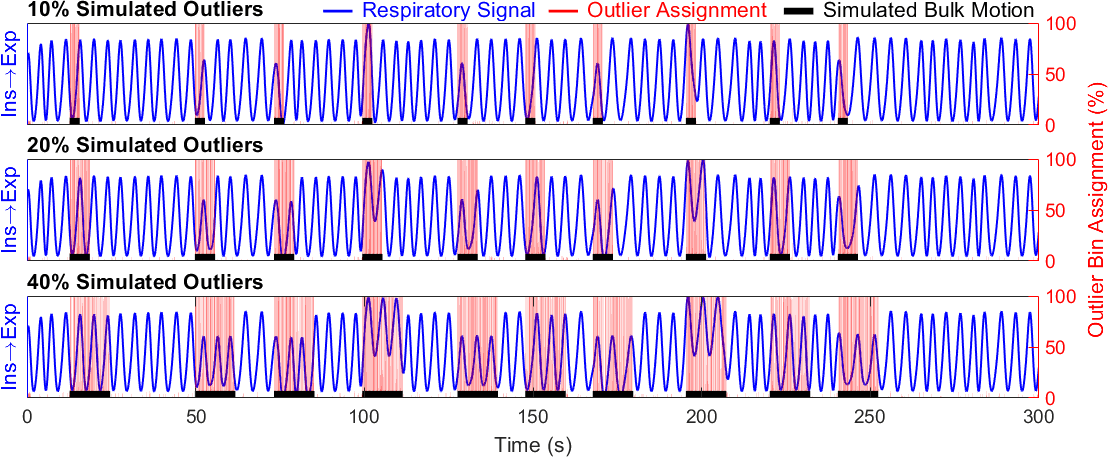}}
\caption{Respiratory surrogate signals (blue) overlaid with simulated bulk motion intervals (horizontal black bars) and the assignment percentage to the outlier bin for the corresponding readout (vertical red bars), shown for three representative cases with 10\%, 20\%, and 40\% motion outlier corruption. The height of each red bar represents the assignment percentage to the outlier bin for the corresponding readout.}
\label{fig:phantom_resp}
\end{figure*}

To evaluate EMORe’s ability to reject corrupted data while preserving valid readouts, Fig.~\ref{fig:phantom_resp} displays respiratory surrogate signals (blue) overlaid with simulated bulk motion intervals (horizontal black bars) and the assignment percentage to the outlier bin for the corresponding readout (vertical red bars). Examples are shown for $10\%$, $20\%$, and $40\%$ outlier scenarios, corresponding to the reconstructions in Fig.~\ref{fig:phantom_fig}.

\subsection{5D MRI in vivo study}
The edge sharpness and the blind artifact scores comparisons between EMORe and CS for the 13 in vivo scans, including three scans with instructed coughing in the last 30 seconds, are summarized in Table~\ref{tab:invivo}. EMORe demonstrated significantly improved ($p<0.05$) blood–myocardium edge sharpness compared to CS. Additionally, blind artifact scoring by two independent expert readers showed consistent and statistically significant improvements for EMORe relative to CS in perceived noise and artifact levels, using paired $t$-test ($p<0.05$) with Bonferroni correction.

\begin{figure*}[!t]
\centerline{\includegraphics[width=5in]{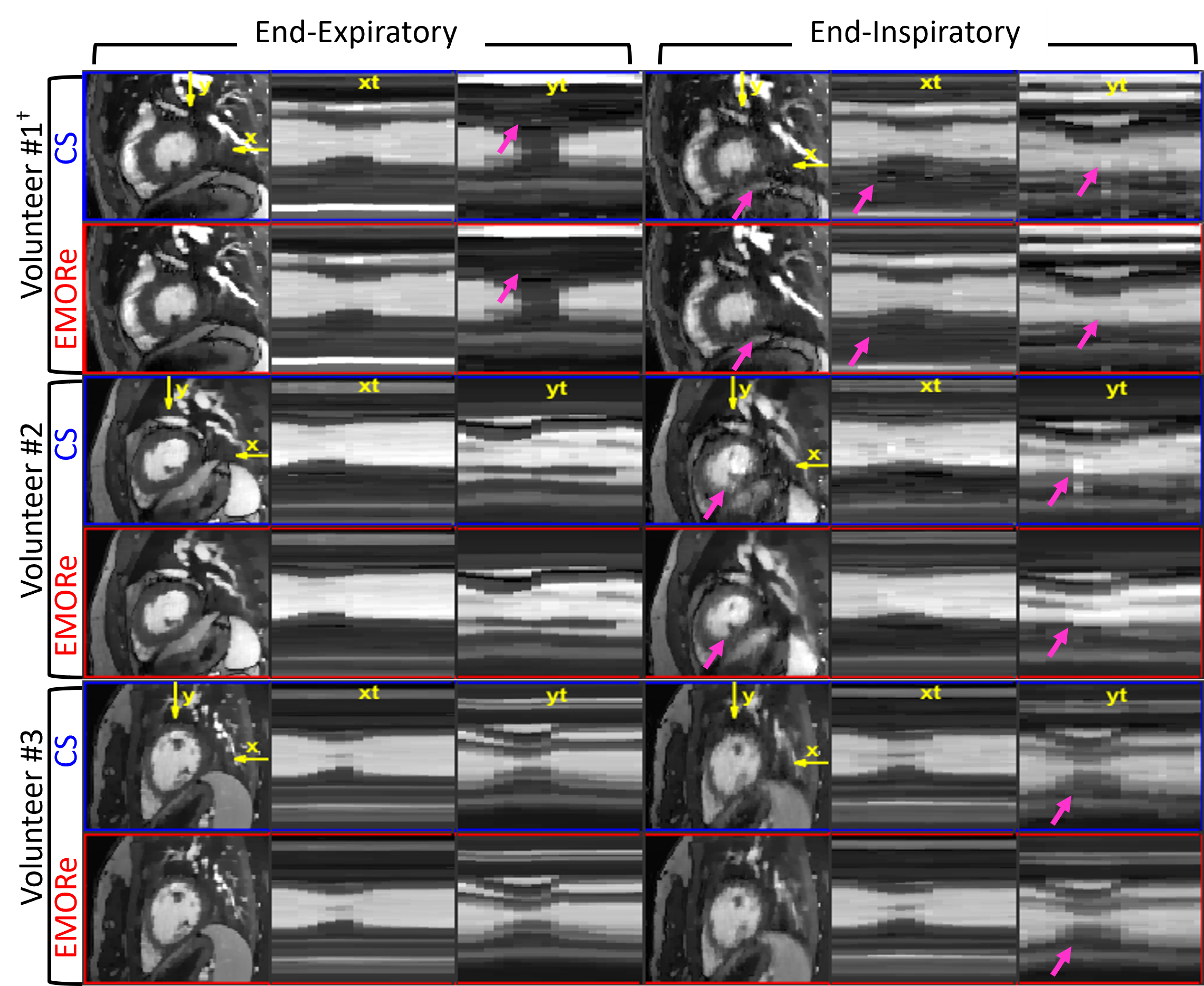}}
\caption{Representative sagittal cine frames at end-expiratory and end-inspiratory phases for three in vivo volunteers reconstructed using CS and EMORe. Corresponding temporal profiles along the $x$–$t$ and $y$–$t$ dimensions are presented for qualitative assessment of temporal consistency and boundary definition. Arrows highlight comparisons of observed differences between EMORe and CS reconstructions.\\
\textsuperscript{†}Volunteer instructed to simulate coughing during the final 30 seconds of the scan.}
\label{fig:real_fig}
\end{figure*}

\begin{figure*}[!t]
\centerline{\includegraphics[width=\textwidth]{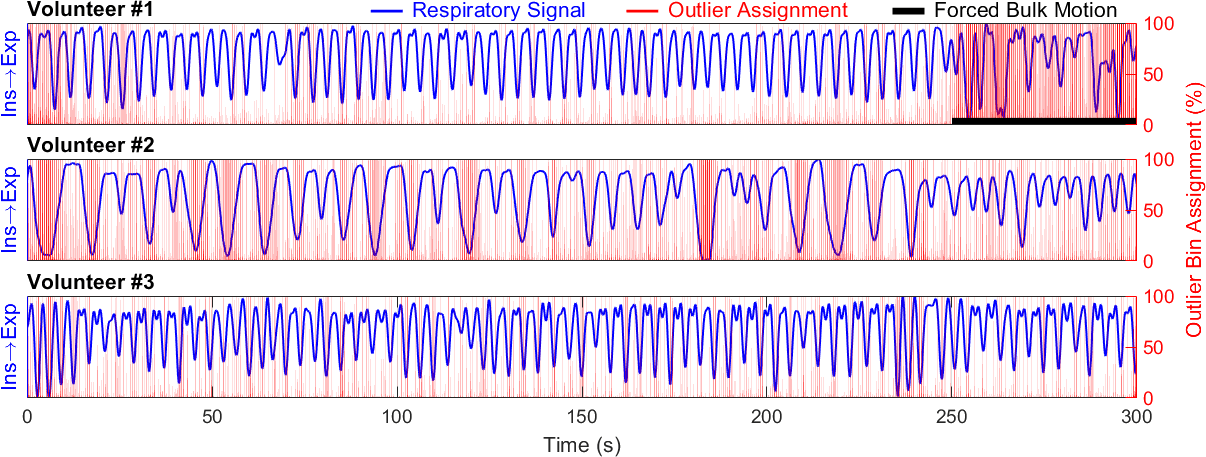}}
\caption{Respiratory surrogate signals (blue) from the three volunteers in Fig.~\ref{fig:real_fig} overlaid with the assignment percentage to the outlier bin for the corresponding readout (vertical red bars). The instructed coughing interval of Volunteer \#1 indicated with (horizontal black bar).}
\label{fig:real_resp}
\end{figure*}

Fig.~\ref{fig:real_fig} presents sagittal cardiac frames and the corresponding horizontal and vertical temporal profiles at end-expiratory and end-inspiratory phases for EMORe and CS reconstructions from three representative volunteer scans. Volunteer \#1 was instructed to simulate coughing motion during the final $30$ seconds of the $5$-minute scan, whereas Volunteers \#2 and \#3 underwent standard free-breathing acquisitions without specific instructions. A video\footref{fn:video} containing all the cine slices presented in Fig.~\ref{fig:real_fig} is available under the “Supplementary Files”. Fig.~\ref{fig:real_resp} presents the respiratory surrogate signals (blue) for the scans corresponding to Fig.~\ref{fig:real_fig}; overlaid with the coughing interval (horizontal black bars) and the outlier bin assignment percentage of the corresponding data (vertical red bars).


\begin{table}
\caption{Comparison of EMORe and CS for edge sharpness and blind image scores (per reviewer) for in vivo study}
\centering
\label{tab:invivo}
\setlength{\tabcolsep}{4pt}
\renewcommand{\arraystretch}{1.2}
\begin{tabular}{|p{0.3\columnwidth}|c|c|}
\hline
\textbf{Metric} & \multicolumn{1}{c|}{\textbf{EMORe}}  & \multicolumn{1}{c|}{\textbf{CS}}\\
& \multicolumn{1}{c|}{\textbf{(Mean ± SEM)}}  & \multicolumn{1}{c|}{(Mean ± SEM)}\\
\hline
Edge sharpness ($\uparrow$) & $0.722 \pm 0.028^*$ & $0.694 \pm 0.023$ \\
\hline
Blind artifact score Reviewer 1 ($\uparrow$) & $4.00 \pm 0.06^*$ & $3.58 \pm 0.13$ \\
\hline
Blind artifact score Reviewer 2 ($\uparrow$) & $4.12 \pm 0.14^*$ & $3.65 \pm 0.19$ \\
\hline
\multicolumn{3}{p{0.91\columnwidth}}{SEM: standard error of the mean.}\\
\multicolumn{3}{p{0.91\columnwidth}}{*Statistically significant improvement as determined by paired $t$-test ($p < 0.05$).}
\end{tabular}
\end{table}

\section{Discussion}
In the phantom experiments, EMORe image quality was indistinguishable from CS in the absence of motion outliers, as measured by PSNR, SSIM, and edge sharpness. Yet even without bulk motion outliers, EMORe achieved statistically significant improvement in the Brier score by adaptively refining a soft binning; this highlights that some inaccuracies in SG-based binning may occur without bulk motion, due to irregular breathing and beat-to-beat variations.
With increasing simulated bulk motion outliers, CS reconstructions exhibited significant degradation, whereas EMORe maintained higher fidelity, reflecting its robustness against sporadic motion artifacts. In extreme cases with more than $40\%$ motion outliers, although EMORe maintained an advantage over CS, the overall image quality of EMORe also degrades substantially, highlighting the limitations of the proposed method under widespread data corruption.

Visual comparisons in Figs.~\ref{fig:phantom_fig} further support these quantitative findings in the phantom study, showing that EMORe reconstructions exhibit clearer anatomical boundaries and reduced motion artifacts relative to CS, particularly at higher outlier levels. Moreover, the respiratory surrogate plots in Fig.~\ref{fig:phantom_resp} demonstrate EMORe’s capability to aggressively reject outlier data during simulated bulk motion instances, resulting in improved Brier scores and preserving the fidelity of data in the valid motion state bins throughout the scan.

From the in vivo study, EMORe reconstructions showed statistically significant improvement
in blood–myocardium edge sharpness, compared to CS recovery without adaptive refinement of bin assignments. Likewise, the expert reviewers' $1-5$ Likert scores, averaged across cine image pairs ($N=26$), improved by $0.45$ via EMORe processing, showing enhanced robustness to motion artifacts. Together, these improvements suggest practical value in clinical application.

Among the $13$ volunteers, $3$ were instructed to perform intermittent coughing to provide a controlled evaluation of EMORe's ability to identify and discard data readouts corrupted by sporadic bulk motion. Results in Fig.~\ref{fig:real_resp} showed aggressive rejection of corrupted readouts during the known coughing episode. For Volunteer \#2, an irregular breathing pattern, interspersed with deep breaths, leads to more frequent rejection of the data into the outlier bin.  

The adaptive binning in the EMORe framework is achieved at a modest computational overhead, as reflected in the reconstruction time increases of approximately $24\%$ and $16\%$ for phantom and in vivo datasets, respectively. Moreover, EMORe required the selection of additional hyperparameters, including the prior weights $\alpha_g$ and $\alpha_o$, as well as the outlier rejection threshold $\tau$, to balance the trade-offs among artifact suppression, outlier rejection, and algorithm stability.

Future directions include replacing the M-step in EMORe with a deep learning-based reconstruction method to potentially further improve the image quality. 
Additionally, extending EMORe to in vivo 5D flow studies could enhance quantitative flow analysis and broaden its clinical applicability. Furthermore, evaluating EMORe in patient populations with irregular breathing, arrhythmias, or fidgeting movement would provide further validation of its robustness in clinical settings.


\section{Conclusion}
EMORe enables motion-robust imaging in 
free-running, free-breathing self-gated 5D cardiac MRI.
The robustness derives from adaptive, probabilistic, retrospective data binning and outlier rejection
implmented in an expectation-maximization framework. 
The EMORe approach adaptively mitigates binning inaccuracies caused by either motion estimation techniques or sporadic bulk patient motion.
For 3D imaging resolved in both cardiac and respiratory phases, the technique offers improved image sharpness and motion artifact suppression, compared to conventional methods.

\appendices

\section*{Acknowledgment}
We acknowledge the Ohio Supercomputer Center for providing the computational resources used in this work. We thank Juliet Varghese for feedback.

\bibliographystyle{ieeetr}
\bibliography{main}

\begin{thebibliography}{10}

\bibitem{dyverfeldt20154d}
P.~Dyverfeldt {\em et~al.}, ``{4D} flow cardiovascular magnetic resonance consensus statement,'' {\em J. Cardiovasc. Magn. Reson.}, vol.~17, no.~1, p.~72, 2015.

\bibitem{salerno2017recent}
M.~Salerno, B.~Sharif, H.~Arheden, A.~Kumar, L.~Axel, D.~Li, and S.~Neubauer, ``Recent advances in cardiovascular magnetic resonance: techniques and applications,'' {\em Circ. Cardiovasc. Imaging}, vol.~10, no.~6, p.~e003951, 2017.

\bibitem{goo2018nav}
H.~W. Goo, ``Comparison between three-dimensional navigator-gated whole-heart {MRI} and two-dimensional cine {MRI} in quantifying ventricular volumes,'' {\em Korean J. Radiol.}, vol.~19, no.~4, pp.~704--714, 2018.

\bibitem{zaitsev2015motion}
M.~Zaitsev, J.~Maclaren, and M.~Herbst, ``Motion artifacts in {MRI}: A complex problem with many partial solutions,'' {\em J. Magn. Reson. Imaging}, vol.~42, no.~4, pp.~887--901, 2015.

\bibitem{yuan2000bellow}
Q.~Yuan {\em et~al.}, ``Cardiac-respiratory gating method for magnetic resonance imaging of the heart,'' {\em Magn. Reson. Med.}, vol.~43, no.~2, pp.~314--318, 2000.

\bibitem{santelli2011nobellow}
C.~Santelli {\em et~al.}, ``Respiratory bellows revisited for motion compensation: preliminary experience for cardiovascular {MR},'' {\em Magn. Reson. Med.}, vol.~65, no.~4, pp.~1097--1102, 2011.

\bibitem{feng2016sg}
L.~Feng, L.~Axel, H.~Chandarana, K.~T. Block, D.~K. Sodickson, and R.~Otazo, ``{XD-GRASP}: golden-angle radial {MRI} with reconstruction of extra motion-state dimensions using compressed sensing,'' {\em Magn. Reson. Med.}, vol.~75, no.~2, pp.~775--788, 2016.

\bibitem{di2019sg}
L.~Di~Sopra, D.~Piccini, S.~Coppo, M.~Stuber, and J.~Yerly, ``An automated approach to fully self-gated free-running cardiac and respiratory motion-resolved {5D} whole-heart {MRI},'' {\em Magn. Reson. Med.}, vol.~82, no.~6, pp.~2118--2132, 2019.

\bibitem{ma2020sg}
L.~E. Ma {\em et~al.}, ``{5D} flow {MRI}: a fully self-gated, free-running framework for cardiac and respiratory motion--resolved {3D} hemodynamics,'' {\em Radiol.: Cardiothorac. Imaging}, vol.~2, no.~6, p.~e200219, 2020.

\bibitem{kustner2021sg}
T.~K{\"u}stner {\em et~al.}, ``Fully self-gated free-running {3D} cartesian cardiac cine with isotropic whole-heart coverage in less than 2 min,'' {\em NMR Biomed.}, vol.~34, no.~1, p.~e4409, 2021.

\bibitem{4dflow_pruitt2021}
A.~Pruitt {\em et~al.}, ``Fully self-gated whole-heart {4D} flow imaging from a 5-minute scan,'' {\em Magn. Reson. Med.}, vol.~85, no.~3, pp.~1222--1236, 2021.

\bibitem{holtackers2024flex}
R.~J. Holtackers and M.~Stuber, ``Free-running cardiac and respiratory motion-resolved imaging: A paradigm shift for managing motion in cardiac {MRI}?,'' {\em Diagnostics}, vol.~14, no.~17, p.~1946, 2024.

\bibitem{larson2004sg_orig}
A.~C. Larson, R.~D. White, G.~Laub, E.~R. McVeigh, D.~Li, and O.~P. Simonetti, ``Self-gated cardiac cine {MRI},'' {\em Magn. Reson. Med.}, vol.~51, no.~1, pp.~93--102, 2004.

\bibitem{feng20185d}
L.~Feng {\em et~al.}, ``{5D} whole-heart sparse {MRI},'' {\em Magn. Reson. Med.}, vol.~79, no.~2, pp.~826--838, 2018.

\bibitem{liu20103dcine}
J.~Liu, P.~Spincemaille, N.~C. Codella, T.~D. Nguyen, M.~R. Prince, and Y.~Wang, ``Respiratory and cardiac self-gated free-breathing cardiac cine imaging with multi-echo {3D} hybrid radial {SSFP} acquisition,'' {\em Magn. Reson. Med.}, vol.~63, no.~5, pp.~1230--1237, 2010.

\bibitem{usman20173dcine}
M.~Usman, B.~Ruijsink, M.~Nazir, G.~Cruz, and C.~Prieto, ``Free breathing whole-heart {3D} cine {MRI} with self-gated {C}artesian trajectory,'' {\em Magn. Reson. Imaging}, vol.~38, pp.~129--137, 2017.

\bibitem{moghari20183dcine}
M.~H. Moghari, A.~Barthur, M.~E. Amaral, T.~Geva, and A.~J. Powell, ``Free-breathing whole-heart {3D} cine magnetic resonance imaging with prospective respiratory motion compensation,'' {\em Magn. Reson. Med.}, vol.~80, no.~1, pp.~181--189, 2018.

\bibitem{stehning2005cor3dcine}
C.~Stehning, P.~B{\"o}rnert, K.~Nehrke, H.~Eggers, and M.~Stuber, ``Free-breathing whole-heart coronary {MRA} with {3D} radial {SSFP} and self-navigated image reconstruction,'' {\em Magn. Reson. Med.}, vol.~54, no.~2, pp.~476--480, 2005.

\bibitem{bhat2011cor3dcine}
H.~Bhat, L.~Ge, S.~Nielles-Vallespin, S.~Zuehlsdorff, and D.~Li, ``{3D} radial sampling and {3D} affine transform-based respiratory motion correction technique for free-breathing whole-heart coronary {MRA} with 100\% imaging efficiency,'' {\em Magn. Reson. Med.}, vol.~65, no.~5, pp.~1269--1277, 2011.

\bibitem{ingle2014cor3dcine}
R.~R. Ingle {\em et~al.}, ``Nonrigid autofocus motion correction for coronary {MR} angiography with a {3D} cones trajectory,'' {\em Magn. Reson. Med.}, vol.~72, no.~2, pp.~347--361, 2014.

\bibitem{wu20125d}
H.~H. Wu, D.~G. Nishimura, B.~S. Hu, and M.~V. McConnell, ``Acquisition and visualization of {5D} respiratory-resolved cardiac {MRI},'' {\em J. Cardiovasc. Magn. Reson.}, vol.~14, no.~Suppl 1, p.~P237, 2012.

\bibitem{sieber20255d}
X.~Sieber {\em et~al.}, ``Ferumoxytol-enhanced free-running {5D} whole-heart {CMR} at 0.55\,{T},'' {\em J. Cardiovasc. Magn. Reson.}, vol.~27, 2025.

\bibitem{weiss2024resp}
E.~K. Weiss {\em et~al.}, ``Respiratory-resolved five-dimensional flow cardiovascular magnetic resonance: In-vivo validation and respiratory-dependent flow changes in healthy volunteers and patients with congenital heart disease,'' {\em J. Cardiovasc. Magn. Reson.}, vol.~26, no.~2, p.~101077, 2024.

\bibitem{feng2016xd}
L.~Feng, L.~Axel, H.~Chandarana, K.~T. Block, D.~K. Sodickson, and R.~Otazo, ``{XD-GRASP}: golden-angle radial {MRI} with reconstruction of extra motion-state dimensions using compressed sensing,'' {\em Magn. Reson. Med.}, vol.~75, no.~2, pp.~775--788, 2016.

\bibitem{arshad2024CORe}
S.~M. Arshad {\em et~al.}, ``Motion-robust free-running volumetric cardiovascular {MRI},'' {\em Magn. Reson. Med.}, vol.~92, no.~3, pp.~1248--1262, 2024.

\bibitem{ludwig2021pt}
J.~Ludwig, P.~Speier, F.~Seifert, T.~Schaeffter, and C.~Kolbitsch, ``Pilot tone--based motion correction for prospective respiratory compensated cardiac cine {MRI},'' {\em Magn. Reson. Med.}, vol.~85, no.~5, pp.~2403--2416, 2021.

\bibitem{roy20245d}
C.~W. Roy {\em et~al.}, ``Intra-bin correction and inter-bin compensation of respiratory motion in free-running five-dimensional whole-heart magnetic resonance imaging,'' {\em J. Cardiovasc. Magn. Reson.}, vol.~26, no.~1, p.~101037, 2024.

\bibitem{kustner2020cinenet}
T.~K{\"u}stner {\em et~al.}, ``{CINENet}: deep learning-based {3D} cardiac cine {MRI} reconstruction with multi-coil complex-valued {4D} spatio-temporal convolutions,'' {\em Sci. Rep.}, vol.~10, no.~1, p.~13710, 2020.

\bibitem{niethammer2007staticoutlier1}
M.~Niethammer, S.~Bouix, S.~Aja-Fern{\'a}ndez, C.-F. Westin, and M.~E. Shenton, ``Outlier rejection for diffusion weighted imaging,'' in {\em International Conference on Medical Image Computing and Computer-Assisted Intervention}, pp.~161--168, Springer, 2007.

\bibitem{dolui2017staticoutlier2}
S.~Dolui, Z.~Wang, R.~T. Shinohara, D.~A. Wolk, J.~A. Detre, and A.~D.~N. Initiative, ``Structural correlation-based outlier rejection ({SCORE}) algorithm for arterial spin labeling time series,'' {\em J. Magn. Reson. Imaging}, vol.~45, no.~6, pp.~1786--1797, 2017.

\bibitem{patel2014staticoutlier3}
A.~X. Patel {\em et~al.}, ``A wavelet method for modeling and despiking motion artifacts from resting-state f{MRI} time series,'' {\em Neuroimage}, vol.~95, pp.~287--304, 2014.

\bibitem{em_mclachlan2008}
G.~J. McLachlan and T.~Krishnan, {\em The {EM} Algorithm and Extensions}.
\newblock John Wiley \& Sons, 2008.

\bibitem{van2018emoutlier}
J.~F. van Amerom {\em et~al.}, ``Fetal cardiac cine imaging using highly accelerated dynamic {MRI} with retrospective motion correction and outlier rejection,'' {\em Magn. Reson. Med.}, vol.~79, no.~1, pp.~327--338, 2018.

\bibitem{arshad2025emore}
S.~M. Arshad, L.~C. Potter, X.~Lei, and R.~Ahmad, ``{EMOR}e: motion-robust {XD-CMR} reconstruction using expectation-maximization ({EM}) algorithm,'' {\em J. Cardiovasc. Magn. Reson.}, vol.~27, 2025.

\bibitem{boyd2011admm}
S.~Boyd {\em et~al.}, ``Distributed optimization and statistical learning via the alternating direction method of multipliers,'' {\em Found. Trends Mach. Le.}, vol.~3, no.~1, pp.~1--122, 2011.

\bibitem{kadir2014high}
S.~N. Kadir, D.~F. Goodman, and K.~D. Harris, ``High-dimensional cluster analysis with the masked {EM} algorithm,'' {\em Neural Comput.}, vol.~26, no.~11, pp.~2379--2394, 2014.

\bibitem{cs_lustig2008}
M.~Lustig, D.~L. Donoho, J.~M. Santos, and J.~M. Pauly, ``Compressed sensing {MRI},'' {\em IEEE Signal Process. Mag.}, vol.~25, no.~2, pp.~72--82, 2008.

\bibitem{wang2004ssim}
Z.~Wang, A.~C. Bovik, H.~R. Sheikh, and E.~P. Simoncelli, ``Image quality assessment: from error visibility to structural similarity,'' {\em IEEE Trans. Image Process.}, vol.~13, no.~4, pp.~600--612, 2004.

\bibitem{edge_ahmad2015}
R.~Ahmad, Y.~Ding, and O.~P. Simonetti, ``Edge sharpness assessment by parametric modeling: application to magnetic resonance imaging,'' {\em Concepts Magn. Reson., Part A}, vol.~44, no.~3, pp.~138--149, 2015.

\bibitem{brier1950brier}
G.~W. Brier, ``Verification of forecasts expressed in terms of probability,'' {\em Mon. Weather Rev}, vol.~78, no.~1, pp.~1--3, 1950.

\bibitem{mrxcat_wissmann2014}
L.~Wissmann, C.~Santelli, W.~P. Segars, and S.~Kozerke, ``{MRXCAT}: Realistic numerical phantoms for cardiovascular magnetic resonance,'' {\em J. Cardiovasc. Magn. Reson.}, vol.~16, pp.~1--11, 2014.

\bibitem{joshi2022sampling}
M.~Joshi, A.~Pruitt, C.~Chen, Y.~Liu, and R.~Ahmad, ``Technical report (v1. 0)--pseudo-random {C}artesian sampling for dynamic {MRI},'' {\em arXiv preprint arXiv:2206.03630}, 2022.

\bibitem{chen2024cardiac}
C.~Chen {\em et~al.}, ``Cardiac and respiratory motion extraction for {MRI} using {P}ilot {T}one--a patient study,'' {\em Int. J. Card. Imaging}, vol.~40, no.~1, pp.~93--105, 2024.

\bibitem{coilbuehrer2007}
M.~Buehrer, K.~P. Pruessmann, P.~Boesiger, and S.~Kozerke, ``Array compression for {MRI} with large coil arrays,'' {\em Magn. Reson. Med.}, vol.~57, no.~6, pp.~1131--1139, 2007.

\bibitem{mapwalsh2000}
D.~O. Walsh, A.~F. Gmitro, and M.~W. Marcellin, ``Adaptive reconstruction of phased array {MR} imagery,'' {\em Magn. Reson. Med.}, vol.~43, pp.~682--690, May 2000.

\end{thebibliography}

\end{document}